# Spatial distribution clamping of discrete spatial solitons due to three photon absorption in AlGaAs waveguide arrays

Darren D. Hudson[1,2], J. Nathan Kutz[3], Thomas R. Schibli[1,2], Demetrios N. Christodoulides[4], Roberto Morandotti[5], and Steven T. Cundiff[1,2,*]

[1] JILA, National Institute of Standards and Technology and University of Colorado, Boulder, CO 80309, USA
[2] Department of Physics, University of Colorado, Boulder, CO 80309, USA
[3] Department of Applied Mathematics, University of Washington, Seattle, WA 20036, USA
[4] College of Optics and Photonics, CREOL, University of Central Florida, Orlando, FL 32816, USA
[5] Institut National de la Recherche Scientifique, Université du Quebec, Varennes Canada
[*] cundiff@jila.colorado.edu

**Abstract:** We observe clamping of the output spatial light distribution of a waveguide array. Using a chirped pulse amplifier we reach peak intensities in the waveguides of ~24 GW/cm$^2$. At this level, three photon absorption in the AlGaAs material clamps the discrete spatial soliton to a set distribution. Further increase in intensity does not change the distribution.

## 1. Introduction

Discrete spatial solitons in AlGaAs waveguide arrays have been intensely researched during the last 20 years. Beginning with their theoretical prediction in 1988 [1], and the first experimental observation in 1998 [2], it has been shown that discrete spatial solitons fundamentally differ from their continuum counterparts [3,4]. For example, discrete solitons can be either bright or dark, a feature resulting from the ability of the waveguide array to exhibit either normal or anomalous diffraction [3]. These exotic properties have made discrete spatial solitons a current topic of research and a potential solution for a wide range of applications [5].

Waveguide arrays have been investigated for use as a mode-locking device [6]. Mode-locking relies on the spatio-temporal discrete self-focusing that occurs when the input pulse has sufficient peak intensity to change its propagation constant in the waveguide [7]. A more complete understanding of the effects of multi-photon absorption, which can be deleterious for mode-locking, is needed as the intra-cavity pulses can have large peak powers when they enter the waveguide array. In this work, we analyze the effects on the spatial distribution of light at the output of the waveguide array as the input peak power is increased well beyond the self-focusing threshold. In this regime, three photon absorption is non-negligible and effectively clamps the spatial distribution, with further increases in peak power having no effect on the distribution.

AlGaAs offers a high nonlinear refractive index coefficient ($n_2 \sim 10^{-13}$ cm$^2$/W) and very precise fabrication methods for growing the wafer and for patterning the waveguides using photolithography. By controlling the alloy concentrations, the band gap can be engineered. The ability to vary the band gap energy allows multi-photon absorption to be minimized at a given wavelength [8]. In particular, certain alloy concentrations of AlGaAs can yield a half band gap energy in the telecommunications C band (1530 to 1565 nm). This condition makes the waveguide array a candidate for use in a mode-locked Erbium doped fiber laser operating at a central wavelength of 1550nm. Although it has been shown that band gap engineering can eliminate the effects of two-photon absorption (2PA) and minimize the effects of three-photon absorption (3PA), at sufficiently high intensities multi-photon absorption invariably becomes non-negligible. Using a conventional Erbium doped mode-locked fiber laser in conjunction with a chirped-pulse amplifier (CPA), we were able to reach intensities in the 24 GW/cm$^2$ range, where 3PA is important.

## 2. Theory

The dynamic exchange of electromagnetic energy in waveguiding systems has long been considered in the context of coupled mode theory [6]. Coupled-mode theory provides an ideal and highly-simplified analytic framework describing the coupling and propagation of electromagnetic energy in waveguides and waveguide arrays, even when subject to the cubic Kerr nonlinearity [1]. The theory assumes that the electromagnetic field is localized transversely in each waveguide and that the exchange of energy between the waveguides can be accurately modeled by an evanescent, linear coupling determined by an overlap integral between waveguides [6]. The nonlinear theory, which includes self-phase modulation, agrees remarkably well with experiments with both CW [4,8] and pulsed femtosecond [9] light. In the waveguide array configuration considered here, nearest-neighbor interactions dominate the waveguide array dynamics. This situation leads to a set of discretely coupled nonlinear Schrodinger equations governing the evolution of electromagnetic energy

$$i\frac{\partial A_n}{\partial z} - \frac{\beta''}{2}\frac{\partial^2 A_n}{\partial t^2} + \gamma |A_n|^2 A_n + i\sigma |A_n|^4 A_n + c(A_{n+1} + A_{n-1}) = 0 \qquad (1)$$

where $A_n$ represents the normalized electric field amplitude in the $n^{th}$ waveguide ($n = -N, ..., -1, 0, 1, ..., N$ for $2N + 1$ waveguides). The linear coupling coefficient is determined experimentally to be $c = 0.82$ mm$^{-1}$ and the nonlinear parameter to be $\gamma = 3.6$ m$^{-1}$ W$^{-1}$. Unlike previous CW experiments [4,7], the femtosecond pulse propagation considered here requires the explicit inclusion of chromatic dispersion. Thus the parameter $\beta_2 = 1.25$ ps$^2$/m is the experimentally measured chromatic dispersion in the waveguide array [8]. The parameter $\sigma$ is the strength of the three-photon absorption. The value of $\sigma$ is not known *a priori* for the AlGaAs waveguides. However, it can be empirically found by matching the simulations to experiments.

For comparison to the experimental observations, simulations of the mode-coupling dynamics as given by the governing equation (Eq. 1) are performed with a pseudo-spectral method that spectrally transforms the time-domain solution and uses a fourth-order Runge-Kutta method for propagation in the waveguide. For all simulations, 41 ($N = 20$) waveguides are considered, which is consistent with our waveguide array geometry and design. To study the effects of chirp on femtosecond pulses, an initial 200 femtosecond hyperbolic secant pulse is launched into the center waveguide ($n = 0$) with either normal, anomalous or no chirp. Thus the initial condition for (Eq 1.) is

$$A_0(0,t) = \eta \operatorname{sech}(\omega t) e^{i\alpha t^2} \qquad (2)$$

where α determines the amount of normal ($α > 0$) or anomalous ($α < 0$) chirp initially on the pulse. The pulse amplitude, $\eta$, and width, $\omega$, are chosen to generate 200 femtosecond pulses for $\alpha = 0$ with peak powers matching experimental conditions. We note that the launch conditions are critical to our conclusions.

## 3. Experimental Setup

To generate the input pulses, we use a mode-locked, Erbium doped fiber laser operating at 1550nm with a repetition rate of 25 MHz and a chirped-pulse amplifier/compressor system (see Fig. 1). Using dispersion compensating fiber (DCF), the normally chirped pulses from the fiber laser are further broadened to several picoseconds to avoid nonlinearities in the amplifier. These stretched pulses are coupled to a bi-directionally pumped Erbium amplifier, which increases the pulse energy by a factor of 7, while maintaining the original pulse shape. The output of the amplifier is temporally compressed/stretched in free-space by a diffraction grating compressor. This system produces pulse energies of 3.5 nJ and allows the pulse chirp to be varied from normal through zero to anomalous. The pulses are characterized by measuring the intensity autocorrelation and the sign of the chirp was verified using frequency resolved optical gating.

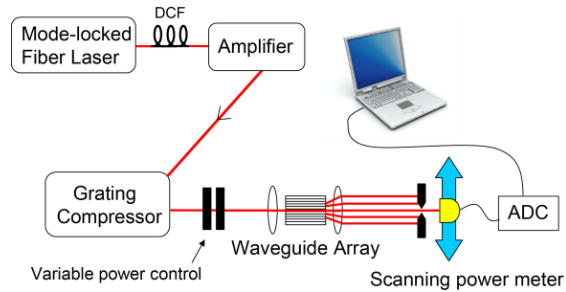

Fig. 1 Experimental Setup. An Erbium-doped mode-locked fiber laser is amplified. A grating compressor re-compresses the pulses. A microscope objective lens couples the pulses into the central waveguide. The output facet is imaged and a slit is scanned through the magnified image to measure the spatial distribution.

The pulses are coupled into the waveguide array using standard microscope objectives (40x) mounted on 3-axis stages. The input field is mode matched to the waveguide with a coupling efficiency > 40%. The waveguide array has a 10 μm center-to-center spacing between waveguides, with 1.5 μm tall ridges and 4 μm wide waveguides. Index guiding in the vertical direction is provided by a core layer consisting of $Al_{0.18}Ga_{.82}As$ and cladding layers of $Al_{0.24}Ga_{.76}As$. The objective lens focuses the input beam to around 20 μm², which is roughly equal to the area of one waveguide mode. This tight focus, together with the high peak power associated with ultrashort pulses, yields a peak intensity of up to 24 GW/cm².

The output spatial distribution is measured by forming a magnified image of the output facet and spatially scanning a slit through the image plane. The intensity transmitted through the slit is recorded. Each waveguide in the array results in a peak in the spatial pattern. The output powers shown in Figs. 2-4 are the peak heights.

## 4. Results

This apparatus was used to record the output spatial power distribution as a function of input power for input pulses with normal, zero and anomalous chirp. The results for normal chirp are shown in Fig. 2. An autocorrelation of the input pulses is shown in Fig. 2(a). The structure on the pulses from the compressor is mainly due to self-phase modulation and higher-order dispersion in the erbium doped amplifier. The output power distribution for this chirp is shown in Fig. 2(b). Self-focusing overcomes discrete diffraction at roughly 7 $GW/cm^2$. Beyond this point, the central waveguide dominates the power distribution as a discrete spatial soliton has formed. As the power is increased further, the spatial soliton should become more localized, i.e., the power in the central waveguide should continue to increase relative to the other waveguides. However this does not appear to be the case, rather the spatial pattern becomes "clamped" to a fixed distribution.

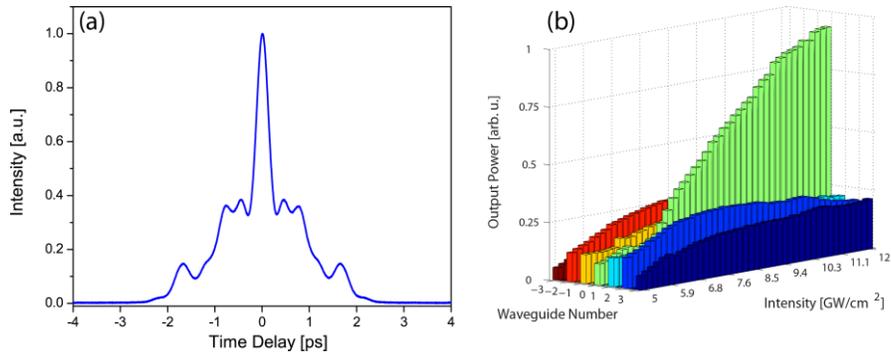

Fig 2. (a) Autocorrelation and (b) power distribution for a normally chirped pulse after the waveguide array. At low input power, the power distribution is spread out and the outer waveguides actually contain more power than the central waveguide. As the input power is increased, discrete diffraction gives way to self-focusing and the central waveguide eventually dominates the power distribution.

To achieve higher intensity levels, the compressor was adjusted to produce the shortest output pulse (i.e., $\alpha \approx 0$). In this case, the shorter pulse (Fig. 3(a)) reaches a maximum peak intensity of 24 $GW/cm^2$. An identical intensity scan (Fig. 3(b)) shows that the clamping of the spatial power distribution is more apparent.

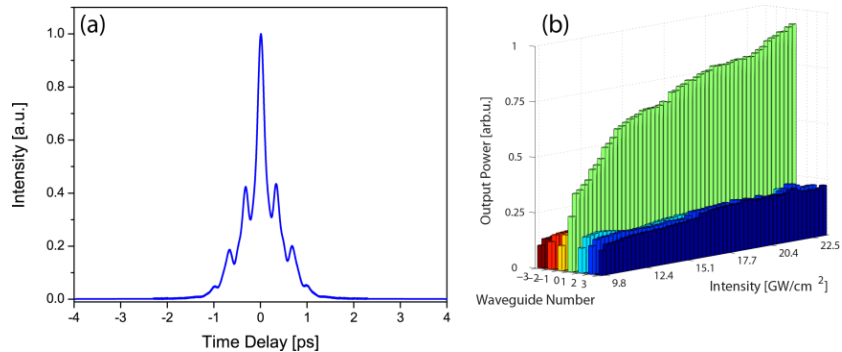

Fig. 3. (a) Autocorrelation and (b) 3D power distribution for an input pulse with zero chirp.

When the compressor is tuned to produce anomalously chirped pulses, with approximately the same duration as the normally chirped pulses used in Fig. 2, the spatial distribution clamps at a slightly lower power (Fig. 4). Since the waveguide array has normal dispersion, input pulses with anomalous chirp re-compress slightly. Thus, anomalously chirped pulses remain shorter inside the waveguide array and thus have an effective higher average peak intensity.

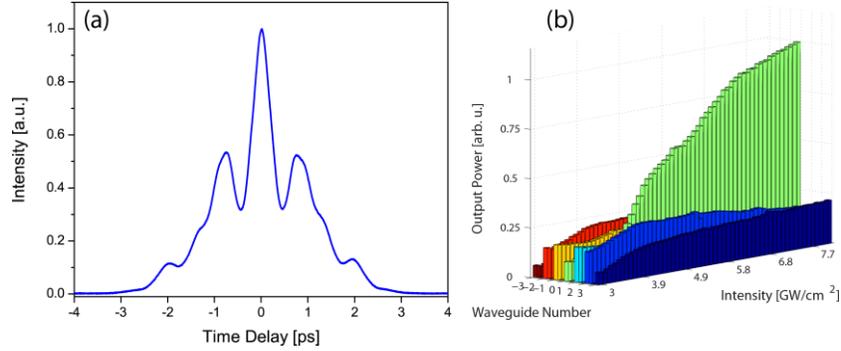

Fig. 4. (a) Autocorrelation and (b) 3D power distribution for an anomalous chirp input pulse.

For the zero chirp case, even as the peak intensity of the input pulse was changed by almost a factor of two, from 14 GW/cm$^2$ to 22.5 GW/cm$^2$, there is only a few percent change in the distribution (within measurement uncertainty). At this high peak power it appears that 3 photon absorption is so strong that it effectively clamps the spatial distribution of power in the array. It is important to note that the absolute power level in each waveguide does not appear to lock, only the relative power between the waveguides.

To make a clear and direct comparison between all of these cases, we plot the ratio of the power in the neighboring waveguides to the power in the central waveguide. Thus, clamping corresponds to the ratio asymptoting to a fixed value. As can be seen in Fig. 5, this limit occurs relatively early in the power scan for the shortest input pulse. The ratio for the anomalously chirped pulse reaches a fixed value at a slightly higher average power than the shortest input pulse, while the ratio for the normally chirped pulse just barely reaches the threshold. By inspection, the intensity at which the distribution settles into a steady state value is around 13 GW/cm$^2$.

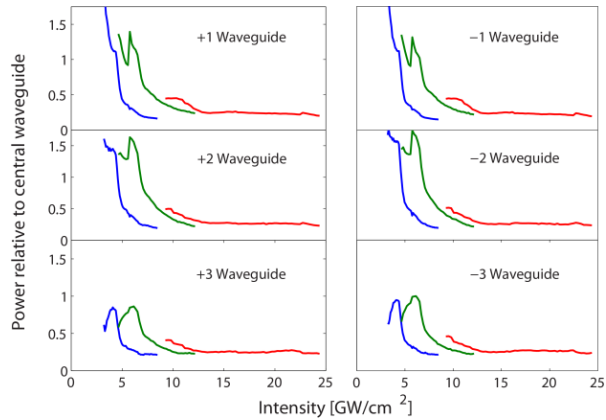

Fig. 5 Relative power in the neighboring waveguides. The shortest input pulse (red) reaches the distribution clamping threshold at around 13 GW/cm2. The anomalous input pulse (green) and normal input pulse (blue) achieve lower peak powers than the shortest pulse due to the limited average power in the setup.

Using Eq. (1) we numerically reproduced the above results, as shown in Fig. 6. The blue line is the relative power with 3PA included in the simulation, while the black line is the relative power without 3PA. These numerical results agree with the experimental data in that the 3PA serves to stabilize the distribution for increasing input intensity. The result of running the simulation with the 3PA term excluded shows that the expected distribution continuously changes as the input intensity is increased, and that the extra energy goes into the central waveguide. The effect of 3PA alters this scenario by balancing this increase in the self-focusing effect with an effective inverse saturable absorber.

As can be seen in the numerical simulation with 3PA included, multi-photon absorption causes the spatial light distribution to settle down to a steady state at intensities around 18 GW/cm$^2$. This is slightly higher than was observed experimentally which implies that the experiment had other inverse saturable absorber contributions that lead to the onset of clamping at lower powers. These contributions could include a larger than expected two-photon absorption due to uncertainties in the alloy concentration or material defects, or perhaps four-photon absorption.

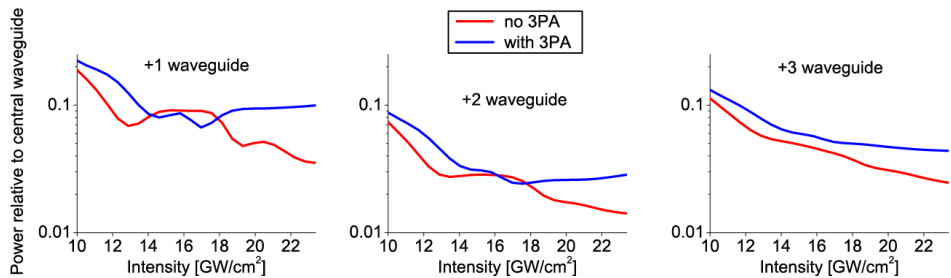

Fig. 6 Numerical simulation of the effect of 3PA on the spatial distribution of the discrete spatial solitons for an input pulse with no chirp. The blue line (3PA term included) converges to a higher relative power than the black line (no 3PA term), and levels off to a steady value. In contrast, the black line monotonically decreases, indicating that the discrete spatial soliton is changing its spatial light distribution by putting more and more energy into the central waveguide.

## 5. Conclusion

In conclusion, the spatial distribution of power in the discrete spatial soliton reaches a steady configuration when the three photon absorption is strong. For chirped pulses, the effect emerges at higher average power since the effect is fundamentally related to the peak power. By looking at three different input chirps and calculating the power ratio, a threshold level becomes apparent, and all three chirps approach it from different slopes. The peak intensity that is needed to reach this threshold is 13 GW/cm$^2$ in the experiment.

This power clamping could find application in realizing a power limiting device to be used just before an amplifier. The large reduction in power fluctuations due to the 3PA in the waveguide array would serve to reduce amplitude noise on the optical field before it is amplified. In a broader sense, this effect could have implications for experiments involving discrete spatial solitons, especially studies in the high intensity regime. These results may also lead to a better understanding of the mode-locking dynamics when the waveguide array is used as a mode-locking device [5]. Presumably, the clamping of this distribution implies a limit on the shortest intra-cavity pulse attainable since the 3PA increases with peak power. Further investigation that combines the measurements presented in this work with the waveguide array mode-locked laser system could help provide a physical understanding of the spatio-temporal dynamics that occurs inside this exciting new laser system.